\documentclass[twocolumn,prb,aps,showpacs,10pt,superscriptaddress]{revtex4-1}
\usepackage{amsfonts}
\usepackage{amsmath}
\usepackage{amssymb}
\usepackage{graphicx}
\newcommand{\mbf}[1]{\mathbf{#1}}
\newcommand{\ddt}[0]{\frac{\partial}{\partial t}}

\renewcommand{\t}[1]{\textrm{#1}}
\renewcommand{\k}[0]{\mbf{k}}

\newcommand{\nn}[0]{\nonumber\\}

\newcommand{\Jsd}[0]{J_{sd}}
\newcommand{\nMn}[0]{n_{Mn}}
\newcommand{\NMn}[0]{N_{Mn}}
\newcommand{\up}[0]{\uparrow}
\newcommand{\down}[0]{\downarrow}
\renewcommand{\Re}[0]{\t{Re}}
\renewcommand{\Im}[0]{\t{Im}}
\begin{document}
\title{Non-perturbative Correlation Effects in Diluted Magnetic Semiconductors}
\author{M.~Cygorek}
\affiliation{Theoretische Physik III, Universit{\"a}t Bayreuth, 95440 Bayreuth, Germany}

\author{P.~I.~Tamborenea}
\affiliation{Departamento de F\'isica and IFIBA, FCEN, Universidad de Buenos Aires, Ciudad
Universitaria, Pabell\'on I, 1428 Ciudad de Buenos Aires, Argentina}
\affiliation{Theoretische Physik III, Universit{\"a}t Bayreuth, 95440 Bayreuth, Germany}

\author{V.~M.~Axt}
\affiliation{Theoretische Physik III, Universit{\"a}t Bayreuth, 95440 Bayreuth, Germany}

\begin{abstract}
The effects of carrier-impurity correlations due to a Kondo-like spin-spin 
interaction in diluted magnetic semiconductors are investigated. These 
correlations are not only responsible for a transfer of spins between the
carriers and the impurities, but also produce non-perturbative effects in the
spin dynamics such as renormalization of the precession frequency of the
carrier spins, which can reach values of several percent in CdMnTe quantum 
wells. 
In two-dimensional systems, the precession frequency renormalization for a 
single electron spin with defined wave vector shows logarithmic divergences
similar to those also known from the Kondo problem in metals. 
For smooth electron distributions, however, the divergences disappear due 
to the integrability of the logarithm. A possible dephasing mechanism 
caused by the wave-vector dependence of the electron spin precession 
frequencies is found to be of minor importance compared to the spin transfer 
from the carrier to the impurity system. In the Markov limit of the 
theory, a quasi-equilibrium expression for the carrier-impurity 
correlation energy can be deduced indicating the formation of
strongly correlated carrier-impurity states for temperatures in the 
mK range.
\end{abstract}
\maketitle
\section{Introduction}
A perturbative treatment of the interaction between quasi-free 
electrons in a metal with localized magnetic impurities predicts logarithmic 
divergences in several quantities such as resistivity and entropy 
at zero temperature\cite{Kondo,Hewson}. 
This finding, the Kondo effect, 
is a famous example of a situation where perturbation theory leads
to unphysical conclusions whereas the measured values of the 
resistivity assume finite values. 
The Kondo problem, i.~e., the question of how to properly describe the 
low-temperature limit of a system with a 
spin dependent carrier-impurity interaction theoretically,
has opened up a wide field of phyiscs.
Although the Kondo problem, as it was originally formulated,
has been solved\cite{Hewson}, the Kondo physics experienced 
a revival since it has become possible
to study experimentally similar problems in other 
systems, e.~g., structures where quantum dots
play the role of the magnetic impurities\cite{KondoDots98,
KondoDots98_2,KondoDotsGlazman99,KondoDots00,
KondoDotsElzerman00,KondoDotsNordlander00,
KondoDots01,KondoSET98,KondoQPC13,KondoSET14}.
The common feature in these systems is
that a microscopic exchange coupling gives rise to an effective
Kondo-Hamiltonian that assumes the form of a spin-spin contact interaction 
between the quasi-free carriers and the localized magnetic impurities, 
or quantum dots, respectively.

Other systems which are usually modelled by a Kondo-like Hamiltonian
are diluted magnetic semiconductors (DMS) where typically II-VI or III-V
semiconductors are doped with magnetic impurities, usually Mn which 
effectively forms a spin-$\frac52$ system. These materials have
been studied extensively in the last decades\cite{Dietl14,WuReview,Cywinski,Nawrocki81,
Crooker97,Wu09,spintronics_dietl,Awschalom07,GaMnAs_synchro15,Perakis2012,
KossutRate3D,FurdynaReview,Krenn,
Awschalom_RSA,Akimov,Roennburg,Perakis_Wang,WuReview,intro,
Camilleri01,FewMn08, HyperfineCdMnTe} 
due to their optical and magnetic properities which make them promising 
candidates for future spintronics devices\cite{DOM,spintronics_ohno,Spintronics}. 
While the carrier-impurity correlations play a crucial role in the 
metallic Kondo problem, they are often neglected in studies about DMS
by employing a mean-field approximation\cite{FurdynaReview}.
In some articles\cite{DasSarma03} it is argued that the situation in DMS is 
different 
from the original Kondo situation in that in the latter only a 
few magnetic impurites and a huge number of quasi-free carriers are present
in the metal, whereas in the former case,
in particular in the case of (intrinsic) II-VI DMS, the number of 
impurities usually exceeds the number of carriers. 

On the other hand, a third-order many-body perturbation 
theory based on the pseudofermion formalism\cite{Morandi10} reveals
Kondo-like divergences in the propagator for the spin dynamics in DMS due 
to the hole-impurity exchange interaction. From this it was concluded that
the carrier-impurity correlations should in fact be important for the 
dynamics in DMS. Furthermore, some 
studies\cite{BMPM12,BMP11,BMP02,BMP02_2,BMP98,BMP96} 
suggest that bound magnetic polarons can play a key role for the 
ferromagnetic behaviour of some DMS. These polarons consist of 
carriers which are bound to the magnetic impurities and, thus,  
the carriers and impurities are strongly correlated.

In this article, we address the question of the importance of 
carrier-impurity correlations in DMS and the relevance of possible
Kondo-like divergences. 
We base our study on a microscopic quantum kinetic theory derived by a 
correlation expansion scheme\cite{Thurn:12} that is capable of 
a non-perturbative description of highly non-equilibrium situations. 
One aspect of the effects of the carrier-impurity correlations on the
spin dynamics has already been found in previous 
works\cite{Thurn:13_1,Thurn:13_2,Cygorek:14_1,PESC, Ungar15}:
The correlations mediate the transfer of spins between 
the carriers and the impurities.
Since in the Markovian limit, the quantum kinetic theory 
contains the special case of rate equations which can also be derived by
a Fermi's golden rule approach\cite{Thurn:13_1}, 
this spin transfer can, in fact, be treated 
perturbatively\footnote{In order to obtain physically meaningful results,
care has to be taken that the mean-field exchange energy between carriers
and impurites is taken into account in the corresponding $\delta$-function
in the expression for Fermi's golden rule\cite{PESC}.}.
Note that in some situations, e.~g., for excitations 
close to the band edge in two- and lower-dimensional DMS\cite{Preceedings},
the Markov limit is not a good approximation so that deviations from
a golden-rule-like exponential decay are predicted.

In the present study, we show that the carrier-impurity correlations 
are also responsible for another effect in the spin dynamics that is not 
predicted by a perturbative method: a renormalization of the 
precession frequency of carrier spins compared with its mean-field value.
It is shown that the frequency renormalization also contains Kondo-like
logarithmic divergences in the Markov limit in two-dimensional systems. 
However, these divergences never lead to unphysical results in the spin
dynamics. 
This is, first of all, due to the fact that the singularities are integrable
and yield finite values for a non-singular spectral electron distribution.
Moreover, the divergence in the frequency renormalization is only 
found for $t\to\infty$ where the amplitude of the precessing 
electron spin has already decayed to zero.
The Markov limit of the quantum kinetic theory also allows to find an 
expression for the carrier-impurity correlation energy which shows a similar
behaviour as the frequency renormalization, including Kondo-like 
logarithmic divergences in the two-dimensional case.

The article is structured as follows:
First, the quantum kinetic theory is briefly reviewed as well as 
effective (PESC, {\it precession of electrons and correlations}) 
equations\cite{PESC} based on the quantum kinetic theory.
Then, the frequency renormalization described by the PESC equations 
is calculated and compared with the result of a Markovian approximation to
the PESC equations in two and three dimensions. A possible electron spin
dephasing mechanism due to the wave vector dependence of the frequency 
renormalization is discussed.
Finally, we investigate the mean carrier-impurity correlation energy.

\section{Theory}
\subsection{System}
The Hamiltonian for conduction band electrons in DMS is modelled by
\begin{subequations}
\begin{align}
&H=H_0+H_{sd}\\
&H_0=\sum_{\k\sigma}\hbar\omega_{\k} c^\dagger_{\sigma\k} c_{\sigma\k}\\
&H_{sd}=\frac\Jsd V \sum_{Inn' \k\k' \sigma\sigma'}
\mbf S_{nn'}\cdot \mbf s_{\sigma\sigma'} 
c^\dagger_{\sigma\k}c_{\sigma'\k'}e^{i(\k'-\k)\mbf R_I} \hat{P}^I_{nn'},
\end{align}
\end{subequations}
where $H_0$ describes the band structure and $H_{sd}$ is the Kondo Hamiltonian
which originates from the exchange interaction between the $s$-type conduction
band electrons and the $d$-electrons of the magnetic ions. Throughout this
article, we assume a parabolic band structure 
with $\omega_\k=\frac{\hbar k^2}{2m^*}$ where $m^*$ is the effective mass.
$\Jsd$ and $V$ are the coupling constant and volume of the DMS,
$c^\dagger_{\sigma\k}$ and $c_{\sigma\k}$ are the creation and annihilation
operators for electrons with spin index $\sigma$ and wave vector $\k$.
$\mbf R_I$ is the position of the $I$-th magnetic impurity and 
$\hat{P}^I_{nn'}=|I,n\rangle\langle I,n'|$ are the projection operators 
corresponding to the spin state of the $I$-th impurity, e.~g., for 
spin $\frac 52$ Mn impurities, $n=\{-\frac 52,-\frac32\dots\frac 52\}$. 
$\mbf S_{nn'}$ and $\mbf s_{\sigma\sigma'}$ are the spin matrices 
for spin $\frac 52$ and $\frac 12$ systems, respectively.

\subsection{Equations of motion}
A microscopic quantum kinetic theory based on a correlation expansion
scheme was constructed in Ref.~\onlinecite{Thurn:12}, where equations of
motion have been derived for the electron and impurity density matrices
$C_{\sigma_1\k}^{\sigma_2}$ and $M_{n_1}^{n_2}$ as well as their correlations
which are defined by
\begin{subequations}
\begin{align}
C_{\sigma_1\k}^{\sigma_2}=&\langle c^\dagger_{\sigma_1\k}c_{\sigma_2\k}\rangle\\
M_{n_1}^{n_2}=&\langle \hat{P}^I_{n_1n_2}\rangle\\
Q_{\sigma_1 n_1\k_1}^{\sigma_2 n_2\k_2}=&V\big(\langle 
c^\dagger_{\sigma_1\k_1}c_{\sigma_2\k_2}e^{i(\k_2-\k_1)\mbf R_I}
\hat{P}^I_{n_1n_2}\rangle +\nn&-
\langle c^\dagger_{\sigma_1\k_1}c_{\sigma_2\k_2}e^{i(\k_2-\k_1)\mbf R_I}
\rangle \langle\hat{P}^I_{n_1n_2}\rangle\big)
\end{align}
\end{subequations}
where the brackets denote the quantum mechanical average as well as an
average over homogeneously distributed impurities.
The equations of motion for these dynamical variables are 
given in Ref.~\onlinecite{Cygorek:14_1}. 

The full quantum kinetic equations are lengthy and their solution requires 
considerable 
numerical effort. However, it was found in Ref.~\onlinecite{PESC} that
they can be drastically simplified in the case 
where the number of impurity ions $\NMn$ is much larger than the 
number of the quasi-free electrons $N_e$. This is usually fulfilled
especially in II-VI DMS where the magnetic doping with Mn does not 
simultaneously lead to p- or n-doping and the carriers stem exclusively
from optical excitation. To understand the effective equations derived in
Ref.~\onlinecite{PESC} it is instructive to first consider the mean-field 
dynamics for the spin $\mbf s_{\k}=\sum_{\sigma_1\sigma_2}
\mbf s_{\sigma_1\sigma_2} C_{\sigma_1\k}^{\sigma_2}$ of electrons 
with wave vector $\k$ and the impurities 
$\langle \mbf S\rangle= \sum_{n_1n_2} \mbf S_{n_1n_2} M_{n_1}^{n_2}$.
In the mean-field approximation, i.~e., if the correlations are neglected, 
one finds
\begin{subequations}
\begin{align}
&\ddt \mbf s_\k\big|_{MF}=\boldsymbol\omega_M \times \mbf s_\k, \\
\label{eq:MFS}&\ddt \langle \mbf S\rangle\big|_{MF}=
-\frac 1{\NMn}\sum_\k \ddt \mbf s_\k\big|_{MF},
\end{align}
\label{eq:MF}
\end{subequations}
where $\boldsymbol\omega_M:=\frac{\Jsd}\hbar\nMn \langle \mbf S\rangle$.
Eq.~(\ref{eq:MFS}) follows from the total spin conservation of the
Kondo Hamiltonian. In the case $\NMn\gg N_e$, the change of the impurity
spin is marginal and can therefore be neglected. The precession of the
electron spin around the mean field due to the impurity magnetization,
on the other hand, is in general important. 
Eq.~(\ref{eq:MF}a) is solved by 
\begin{align}
&\mbf s_\k= R_{\langle \mbf S\rangle}(\omega_M t) \mbf s'_\k,
\end{align}
where $R_{\mbf n}(\alpha)$ is the matrix describing a rotation around 
the vector $\mbf n$ with angle $\alpha$
and the precession frequency 
$\omega_M=\boldsymbol\omega_M\cdot \langle\mbf S\rangle /
|\langle\mbf S\rangle |$ is defined so that it has the same sign as the
coupling constant $\Jsd$. 
In the mean field approximation $\mbf s'_\k$ is constant. However, if
we also account for the carrier-impurity correlations, $\mbf s'_\k$ 
changes slowly with time and constitutes the electron spin 
in a rotating frame.
If the correlations are formally integrated and inserted into the corresponding
equations of motion for the electron variables,
the effective equations for the electron spin component 
${\mbf s'}^\perp_{\k_1}$ perpendicular to
the impurity magnetization can be given as\cite{PESC}:
\begin{align}
\ddt {\mbf s'}^\perp_{\k_1}=&-\sum_\k\bigg[
\Re(G_{\omega_\k}^{\omega_{\k_1}-\omega_M})\big(\frac{b^+}2-b^0n^\up_\k\big)
{\mbf s'}^\perp_{\k_1}+\nn&+
\Re(G_{\omega_\k}^{\omega_{\k_1}+\omega_M})\big(\frac{b^-}2+b^0n^\down_\k\big)
{\mbf s'}^\perp_{\k_1}+\nn&+
\Re(G_{\omega_\k}^{\omega_{\k_1}})\frac{b^\|}2
\big({\mbf s'}^\perp_\k+{\mbf s'}^\perp_{\k_1}\big)\bigg]+\nn&-
\frac{\langle \mbf S\rangle}{|\langle \mbf S\rangle|}\times \sum_\k\bigg[
\Im(G_{\omega_\k}^{\omega_{\k_1}-\omega_M})\big(\frac{b^+}2-b^0n^\up_\k\big)
+\nn&
-\Im(G_{\omega_\k}^{\omega_{\k_1}+\omega_M})\big(\frac{b^-}2+b^0n^\down_\k\big)
\bigg] {\mbf s'}_{\k_1}^\perp,
\label{eq:PESCperp}
\end{align}
The coefficients in Eq.~(\ref{eq:PESCperp}) are given by
$b^{\pm}:=\langle {S^\perp}^2\rangle\pm \frac{\langle S^\|\rangle}2$,
$b^{0}:=\frac{\langle S^\|\rangle}2$
and 
$b^{\|}:={\langle {S^\|}^2 \rangle}$,
where the component of the impurity spin operator in the direction of 
the mean impurity spin is 
$S^\|:= \hat{\mbf S} \cdot \frac{\langle \hat{\mbf S}\rangle}{
|\langle \hat{\mbf S}\rangle|}$,
and the relevant second moments of the impurity spin operator can be 
separated into parallel $\langle {S^\|}^2\rangle$ and
perpendicular parts
$\langle {S^\perp}^2\rangle=\frac 12\langle S^2-{S^\|}^2 \rangle$.
The memory function 
\begin{align}
G_{\omega_\k}^{\omega_{\k_1}}&:=
\frac{\Jsd^2}{\hbar^2}\frac{\nMn}V\int\limits_{-t}^0 dt' 
e^{i(\omega_\k-\omega_{\k_1})t'},
\label{eq:memory}
\end{align}
has to be interpreted as an integral operator and the time-dependent variables 
that appear after $G_{\omega_\k}^{\omega_{\k_1}}$ in Eq.~(\ref{eq:PESCperp})
are evaluated at $t'$.
Finally, $n^{\up/\down}_\k$ are the occupation numbers of the states
with wave-vector $\k$, i.~e., the diagonal elements of the density matrix
with respect to the spin indices.
Eq.~(\ref{eq:PESCperp}) together with the corresponding equations 
for $n^{\up/\down}_{\k}$ given in Ref.~\onlinecite{PESC} are called
{\it precession of electron spins and correlations (PESC)} equations,
since besides the electron spin, also  
the correlations $Q_{\beta\k_1}^{\alpha\k_2}:=\sum_{\sigma_1\sigma_2}
\sum_{n_1n_2}\mbf s_{\sigma_1\sigma_2}\cdot\mbf S_{n_1n_2} 
Q_{\sigma_1n_1\k_1}^{\sigma_2 n_2\k_2}$ 
exhibit a precession-like movement
around the mean field due to the impurity magnetization. 
Note that Eq.~(\ref{eq:PESCperp}) is equivalent to 
the full quantum kinetic theory of Ref.~\onlinecite{Thurn:12} except that
some source terms for the correlations are neglected that
are numerically insignificant [cf. Ref.~\onlinecite{PESC} for details].

Eq.~(\ref{eq:PESCperp}) is only complicated and numerically challenging due
to the time integral induced by the memory function
$G_{\omega_\k}^{\omega_{\k_1}}$. Now, working in the rotating frame 
allows us to assume that the electron variables change only slowly in time
and can equally well be evaluated at $t$ instead of $t'$.
The memory integral consists then only of 
\begin{align}
\int\limits_{-t}^0 dt'\; e^{i(\omega_\k-\omega_{\k_1})t'}
\approx \pi\delta(\omega_\k-\omega_{\k_1})
-i\mathcal{P}\frac 1{\omega_\k-\omega_{\k_1}},
\label{eq:markov}
\end{align}
where $\mathcal{P}$ is the Cauchy principal value.
The Markov approximation (\ref{eq:markov}) was established by 
letting $t\to\infty$ in the
lower limit of the integral and using the Sokhotski-Plemelj theorem.
The validity of the Markovian approximation can in general depend on the
values of $\k$, $\k_1$, $t$ as well as the timescale of the change of the 
electron variables and therefore has to be checked numerically.

If only the real part of the memory function is used in Markov approximation 
and the imaginary part is neglected, the PESC-equations assume a 
golden rule-type form, where the spin transfer dynamics follows approximately 
an exponential decay to the equilibrium value with rate
\begin{align}
\big(\tau_\perp\big)^{-1}\approx& \frac{\Jsd^2\nMn}{\hbar^2 V} \pi 
\Big[D(\omega_1-\omega_M)\frac{b^+}2 +D(\omega_1+\omega_M)\frac{b^-}2 +
\nn&+D(\omega_1)b^\|\Big]
\label{eq:rate}
\end{align}
for an electron with kinetic energy $\omega_1$, if the terms of second 
order of the electron variables in Eq.~(\ref{eq:PESCperp}) are 
neglected\cite{PESC}.  In the expression for the rate, $D(\omega)$ 
describes the spectral density of  states and depends on the 
dimensionality of the system.

\subsection{Frequency renormalization in the Markov limit}
One issue that we would like to focus on in the present work is the 
change in the
precession frequency described in Eq.~(\ref{eq:PESCperp}) by the terms
proportional to the imaginary part of the memory function. Such a
renormalization of the precession frequency would be absent in any
truncated perturbative approach\cite{Doniach}. It originates, like the 
spin transfer described by the real part of the memory function, from
the carrier-impurity correlations. 
 
It is noteworthy that the frequency renormalization is singular in the 
Markov limit described in Eq.~(\ref{eq:markov}), 
i.~e., the imaginary part of the memory function
$G_{\omega_\k}^{\omega_{\k_1}}$ diverges if $\omega_\k=\omega_{\k_1}$.
However, this divergence does not lead to an unphysical behaviour. 
First of all, the divergence is a feature of the Markovian limit. 
For finite times $t$, the l.~h.~s. of Eq.~(\ref{eq:markov}) is a finite 
integral over an analytic function and is therefore also analytic. For 
$\omega_\k=\omega_{\k_1}$, the value of the integral is $t$ which only 
goes to infinity in the Markov limit. As only the electron spin component
perpendicular to the impurity magnetization is affected by the frequency
renormalization and this component decays approximately exponentially to 
zero, an infinite precession frequency is never observable.

Similar to the Markovian spin transfer rate in Eq.~(\ref{eq:rate}), 
an expression for the frequency renormalization $\Delta\omega$
can be given in the Markov limit of Eq.~(\ref{eq:PESCperp}),
if the imaginary part of Eq.~(\ref{eq:markov}) is used:

\begin{align}
&\Delta\omega(\omega_1)=\frac{\Jsd^2}{\hbar^2}\frac{\nMn}V 
\int\limits_0^{\omega_{BZ}}d\omega \;D(\omega) \times\nn&
\bigg[\frac{b^+}2 \frac 1{\omega-(\omega_1-\omega_M)}
-\frac{b^-}2 \frac 1{\omega-(\omega_1+\omega_M)}\bigg],
\label{eq:freqrengen}
\end{align}
where, for the sake of simplicity, the terms proportional to $n^{\up/\down}$
in Eq.~(\ref{eq:PESCperp}) were neglected, 
since they only matter if a large number of carriers is present.  
In two-dimensional systems, the spectral density of states 
$D^{2d}(\omega)=\frac{A m^*}{2\pi\hbar}\Theta(\omega)$ is constant, where 
$A$ is the sample area 
and $\Theta(x)$ is the step function.
In three dimensions, $D^{3d}(\omega)=\frac{V}{4\pi^2}
\Big(\frac{2m^*}\hbar\Big)^{3/2}\sqrt{\omega}\;\Theta(\omega)$ is 
proportional to the square-root of $\omega$. 
The corresponding frequency renormalizations are:
\begin{subequations}
\begin{align}
\label{eq:dw2d}
\Delta\omega^{2d}(\omega_1)=&
-\frac{\Jsd^2}{\hbar^2}\frac{\nMn}d\frac{m^*}{2\pi\hbar}\bigg\{
\frac{b^+}2\ln\bigg|\frac{\omega_1-\omega_M}
{\omega_{BZ}-(\omega_1-\omega_M)}\bigg|+\nn&
-\frac{b^-}2\ln\bigg|\frac{\omega_1+\omega_M}
{\omega_{BZ}-(\omega_1+\omega_M)}\bigg|\bigg\},
\end{align}
where $d=V/A$ is the quantum well width, 
and
\begin{align}
\Delta\omega^{3d}(\omega_1)=&\frac{\Jsd^2}{\hbar^2}\frac{\nMn}{4\pi}
\Big(\frac{2m^*}\hbar\Big)^{3/2} \int\limits_0^{\omega_{BZ}}d\omega \times
\nn&\bigg\{
\frac{b^+}2\frac{\sqrt{\omega}}{\omega-(\omega_1-\omega_M)}
-\frac{b^-}2\frac{\sqrt{\omega}}{\omega-(\omega_1+\omega_M)}
\bigg\},
\label{eq:w3d}
\end{align}
with
\begin{align}
&\int\limits_0^{\omega_{BZ}}d\omega\frac{\sqrt{\omega}}{\omega-\omega_0}=
\nn&=\left\{\begin{array}{ll}
2\sqrt{\omega_{BZ}} -\sqrt{\omega_0}\ln\Big|\frac{\omega_0+\omega_{BZ}}
{\omega_0-\omega_{BZ}}\Big|,& \omega_0>0 \\
2\sqrt{\omega_{BZ}} -2\sqrt{|\omega_0|}
\t{tan}^{-1}\Big({\frac{\omega_{BZ}}{|\omega_0|}}\Big),&\omega_0<0 
\end{array}\right.
\label{eq:int3dresult}
\end{align}
\label{eq:markovRen}
\end{subequations}
It should be noted that in two and three dimensions
the frequency renormalization depends explicitly on the frequency 
$\omega_{BZ}$, which corresponds to the energy at the end of the first
Brillouin zone, and diverges in the limit $\omega_{BZ}\to\infty$. 
For typical pump-probe experiments with diluted magnetic 
semiconductors, carriers are optically excited relatively close to the 
band edge.
For the excited electrons, one can safely assume 
$\omega_1\pm \omega_M\ll \omega_{BZ}$. In this case, we find a 
logarithmic dependence on $\omega_{BZ}$ in the two-dimensional 
frequency renormalization. 

With the same assumption 
also the integral in Eq.~(\ref{eq:int3dresult})
for the three-dimensional renormalization can be simplified to
\begin{align}
&\int\limits_0^{\omega_{BZ}}d\omega\frac{\sqrt{\omega}}{\omega-\omega_0}\approx
2\sqrt{\omega_{BZ}} -\pi\sqrt{|\omega_0|}\Theta(-\omega_0).
\end{align}
Thus, we find a square-root dependence
of the frequency renormalization on the cut-off frequency $\omega_{BZ}$, 
as well as a square-root dependence on $\omega_0=\omega_1\pm\omega_M$
which only contributes if $\omega_0$ is negative.

The divergence in the limit $\omega_{BZ}\to\infty$ 
is similar to the metallic Kondo effect where the divergence in the
resistivity is also logarithmic in the bandwidth\cite{Hewson}. The 
Kondo problem in metals resembles rather the two-dimensional than 
the three-dimensional case in DMS,
because the formalism for the solution of the Kondo problem usually 
describes the carrier system as possessing a constant spectral density of 
states\cite{Hewson}.

It is noteworthy that the divergence of the frequency renormalization
for $\omega_0=\omega_1\pm \omega_M$
vanishes in the three-dimensional case due to the integral over the 
density of states. In two-dimensional systems, a diverging frequency 
renormalization remains, but only for electrons with a unique value of
the kinetic energy. For realistic optical excitation, however, a smooth
spectral electron distribution can be expected so that the
change of the total precession frequency comprises an averaging over 
frequency renormalizations of nearby states. Since the logarithmic divergence
is integrable, the total frequency renormalization remains finite.
\section{Numerical calculations}
In order to check the validity of the Markov approximation for the 
renormalization of the precession frequency of the electrons, we
compare the Markov result with calculations, where the
memory is taken into account explicitly. It seems straightforward to 
use Eq.~(\ref{eq:PESCperp}) with the time-integral operator 
$G_{\omega_{\k}}^{\omega_{\k_1}}$ defined in Eq.~(\ref{eq:memory})
and solve the integro-differential equations numerically. This is, 
however, a very challenging problem for the following reasons.

From the Markovian expression 
for the frequency renormalization, we find the explicit dependence on
the value of the cut-off energy $\hbar\omega_{BZ}$. Therefore, also 
oscillations with frequencies close to $\omega_{BZ}$ have to be resolved,
which are on the timescale of a few fs since $\hbar\omega_{BZ}$ is in 
the eV range. On the other hand, relevant changes of the total electron spin
takes place in the 10-100 ps range. Furthermore, for each time step the 
calculation of each $\mbf s'^\perp_{\k_1}$ requires a sum over all possible
$\k$-states so that the problem has the complexity $\mathcal{O}(N_k^2)$
where $N_k$ is the number of discretization points for the k-space. 
Note that also in $k$-space, the details of excitations close
to the band edge in the meV range as well as the full Brillouin zone up
to energies of a few eV have to be resolved.
Such a problem also arises in the metallic Kondo effect where numerical
procedures, such as the famous renormalization group\cite{Wilson},
have been developed to deal with the large 
value of the band width\cite{Hewson}.
Note that solving the integro-differential equation by finding an auxiliary
variable, so that the problem can be transformed into an ordinary
differential equation, is equivalent to using the original quantum 
kinetic theory\cite{PESC}.

Here, we solve this problem by using  approximations that allow a separation
of electron spins with different wave vectors, so that we find a 
$\mathcal{O}(N_k)$ problem for an individual electron with wave vector $\k_1$.
First of all, it is noteworthy that $\mbf s'^\perp_{\k_1}$ in 
Eq.~(\ref{eq:PESCperp}) couples to the occupations $n^{\up/\down}_\k$ 
of states with different wave vectors $\k$. 
These terms, however, are of second order in electron variables and  
have a marginal effect on the dynamics of the perpendicular 
spin component\cite{PESC}, especially if the electron density is small, as
is usually the case for optically excited carriers. Neglecting these terms,
we can formulate equations of motion for the complex perpendicular electron
spin variable (in the rotating frame):
\begin{align}
&s'_{\k_1}:= s'^x_{\k_1}+ is'^y_{\k_1},
\end{align}
where it is assumed that the impurity magnetization points in the z-direction.
Then, the PESC-equations~(\ref{eq:PESCperp}) assume the form:
\begin{align}
&\ddt s'_{\k_1}(t)=-\frac{\Jsd^2}{\hbar^2}\frac{\nMn}V\sum_\k\int\limits_0^t dt'
\times\nn&\bigg\{ 
\frac{b^+}2 e^{i[\omega_\k-(\omega_{\k_1}-\omega_M)](t'-t)}s'_{\k_1}(t')
+\nn&
+\frac{b^-}2 e^{-i[\omega_\k-(\omega_{\k_1}+\omega_M)](t'-t)}s'_{\k_1}(t')
+\nn&
+\frac{b^\|}2 \cos[(\omega_\k-\omega_{\k_1})(t'-t)]
\big(s'_{\k}(t')+s'_{\k_1}(t')\big)\bigg\}
\label{eq:PESCreform}
\end{align}
It can be seen immediately from Eq.~(\ref{eq:PESCreform}) that 
in the equation for $s'_{\k_1}$, electron variables of states with other
wave vectors only enter in the last term, i.~e., the term proportional
to $s'_{\k}(t')+s'_{\k_1}(t')$. 
Note that a time integration of $\cos[(\omega_\k-\omega_{\k_1})(t'-t)]$
yields $\frac{\sin[(\omega_\k-\omega_{\k_1})t]}{(\omega_\k-\omega_{\k_1})t}$
which has a pronounced peak at $\omega_\k=\omega_{\k_1}$.
Thus, if the electron spin distribution is assumed to be a smooth function 
in k-space, the main contribution of the last term in Eq.~(\ref{eq:PESCreform}) 
will be approximately the same if we set 
\begin{align}
s'_{\k}(t')\approx s'_{\k_1}(t').
\label{eq:decoupleApprox}
\end{align}
This approximation was shown to 
reproduce the non-Markovian features of the spin transfer 
in Ref.~\onlinecite{Preceedings}.
Also, in contrast to the other terms, the last term of 
Eq.~(\ref{eq:PESCreform}), where the approximation is used, 
does not influence the frequency renormalization, due to the
absence of an imaginary part of the oscillating prefactor
$\cos[(\omega_\k-\omega_{\k_1})(t'-t)]$.
Now, with the help of approximation~(\ref{eq:decoupleApprox}), we end
up with completely decoupled equations for the spins $s'_{\k_1}$ 
of electrons with different wave vectors $\k_1$.

Finally, it is useful for the numerical solution of the 
integro-differential equation~(\ref{eq:PESCreform}) to transform it
into an ordinary differential equation using auxiliary variables
$G_{\k_1\k}^j$:
\begin{subequations}
\label{eq:num}
\begin{align}
&\ddt s'_{\k_1}=-\frac{\Jsd^2}{\hbar^2}\frac{\nMn}V\sum_{j=1}^4
\sum_\k D(\k) G_{\k_1\k}^j,\\
&\ddt G_{\k_1\k}^j=\sigma_j i (\omega_\k-\omega_{\k_1}+\chi_j\omega_M)
G_{\k_1\k}^j+\frac{b_j}2 s'_{\k_1},
\end{align}
with 
\begin{align}
&\sigma_j=\{1,-1,1,-1\},\\
&\chi_j=\{1,-1,0,0\},\\
&b_{j}=\{b^+,b^-,b^\|,b^\|\}
\end{align}
\end{subequations}
and initial conditions $G_{\k_1\k}^j=0$ for $t=0$.
Calculating the dynamics of a single electron spin using 
Eqs.~(\ref{eq:num}) has the complexity $\mathcal{O}(N_k)$ 
and can be done without the need for a numerical 
renormalization group procedure.

\section{Results for the Frequency Renormalization}
The parameters used for the numerical calculations describe a 
Cd$_{0.93}$Mn$_{0.07}$Te
sample with coupling constant $\Jsd=-15 $ meVnm$^3$, effective mass
$m^*=0.093$ $m_0$\cite{Cardona}, where $m_0$ is the free electron mass, 
and, in the case
of a two-dimensional system, a quantum well width of $d=5$ nm. The 
cut-off energy was taken to be $\hbar\omega_{BZ}=3$ eV. The initial 
impurity magnetization was modelled to be thermally distributed and 
is therefore completely defined by the mean value  
$\langle S^\|\rangle\in[-\frac 52;\frac 52]$.

\begin{figure}[t]
\includegraphics{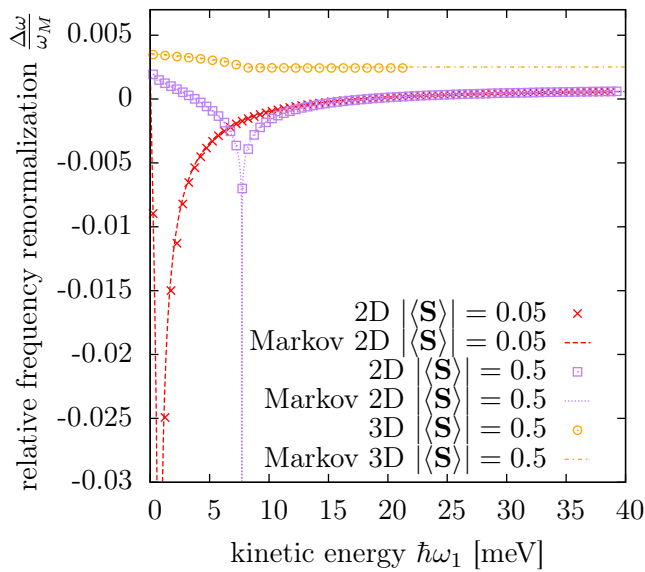}   
\caption{Dependence of the relative frequency renormalization 
$\frac{\Delta\omega}{\omega_m}$ on the kinetic electron energy for
two- and three-dimensional systems according to the calculation 
including a finite memory [Eqs.~(\ref{eq:num})] (points) and in the 
Markov limit [Eqs.~(\ref{eq:markovRen})]  (lines) for different values of the
average impurity spin $|\langle\mbf S\rangle|$ (in units of $\hbar$).  
}
\label{fig:renorm}
\end{figure}

We assume that electrons have been spin selectively prepared by optical
excitation with circularly polarized light so that the initial electron spin
is perpendicular to the initial impurity magnetization (Voigt geometry).
Eqs.~(\ref{eq:num}) are used to calculate the finite-memory 
spin dynamics for electrons with a defined wave vector $k_1$ or, 
equivalently, kinetic energy $\hbar\omega_1=\frac{\hbar^2 k_1^2}{2m^*}$.
An exponentially decaying cosine 
\begin{align}
&s'^x_{\omega_1}(t) \approx s'^x_{\omega_1}(0) e^{-t/\tau_\perp}
\cos(\omega_M' t)
\end{align}
is fit to the non-Markovian spin dynamics in order to find a 
value for the effective decay rate 
$\tau_\perp^{-1}(\omega_1)$ and the precession frequency 
$\omega_M'(\omega_1)$. 
The relative renormalization of the precession frequency is given by 
$\frac{\Delta\omega}{\omega_M}$ with $\Delta\omega=\omega'_M-\omega_M$.

Fig.~\ref{fig:renorm} shows the relative frequency renormalization obtained
from a fit to the non-Markovian calculation and the corresponding 
Markovian result for a $\delta$-like initial spectral electron distribution 
as a function of the kinetic energy $\hbar\omega_1$. First of all,
it can be seen that in three-dimensional as well as in 
two-dimensional systems the Markovian and non-Markovian results coincide.
In the three-dimensional case, the square-root energy dependence of the 
renormalization for $\omega_1<\omega_M$ can be seen clearly, while in 
two dimensions, the logarithmic divergence at $\omega_1=\omega_M$ is 
apparent. The positive relative frequency renormalization in three
dimensions describes an increase in the modulus of the precession frequency. 
In two dimensions, the slightly positive background of the renormalization
is overcompensated by a negative value in the region around the divergence.

\begin{figure}[t]
\includegraphics{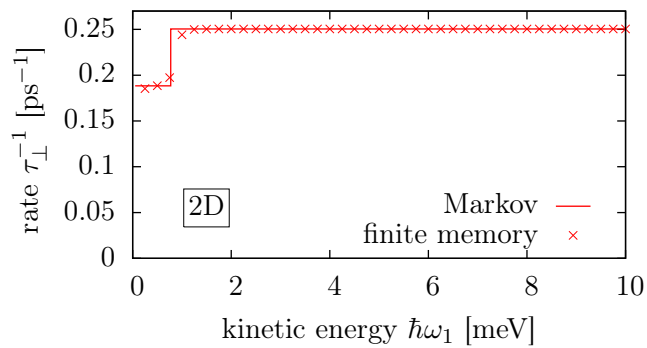}  
\caption{Spin transfer rate $\tau_\perp^{-1}$ 
in Markov approximation (line) according to 
Eq.~(\ref{eq:rate}) and exponential fit to the calculation including a
finite memory (crosses) using Eq.~(\ref{eq:num}) for 
$|\langle\mbf S\rangle|=0.05$ in a two-dimensional system.
}
\label{fig:rate}
\end{figure}

In Fig.~\ref{fig:rate}, the spin transfer rate according to the Markov 
approximation is compared with the value obtained by the exponential 
fit to the non-Markovian result for a calculation with
$|\langle\mbf S\rangle|=0.05$ in two dimensions. The step in the rate
$\tau_\perp^{-1}$ at $\omega_1=\omega_M$, which is predicted in the 
Markov limit [cf. Eq.~(\ref{eq:rate})], is found to be slightly 
rounded off in the non-Markovian calculation, but the deviations 
between both results are rather small.

\begin{figure}[t]
\includegraphics{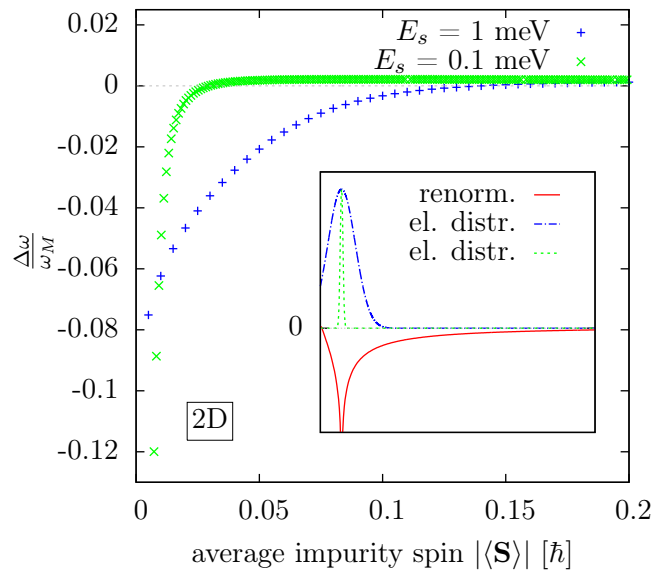}  
\caption{Relative frequency renormalization $\frac{\Delta\omega}{\omega_M}$
in a two-dimensional system
for a Gaussian spectral electron distribution centered at $E_c=\hbar\omega_M$
with standard deviations of $E_s=1$ meV and $E_s=0.1$ meV, respectively. 
The initial electron distribution 
as a function of the kinetic energy is visualized in the inset as the blue
dash-dotted line ($E_s=1$ meV) and green dotted line ($E_s=0.1$ meV) 
together with the corresponding frequency renormalization for 
$\delta$-like excitations (red line) for $|\langle\mbf S\rangle|=0.05$. 
}
\label{fig:gauss}
\end{figure}

In order to find an estimate for the strength of the change of the
precession frequency for a more realistic electron distribution, 
Fig.~\ref{fig:gauss} shows the relative precession frequency renormalization 
as a function of the average impurity spin where the initial spectral 
electron distribution [cf. inset of Fig.~\ref{fig:gauss}]
was assumed to be Gaussian with center at $E_c=\hbar\omega_M$ and 
standard deviation $E_s=1$ meV  ($0.1$ meV) 
corresponding to a full width at half
maximum (FWHM) of $\approx 2.35$ meV ($0.235$ meV) 
or a Gaussian envelope of an exciting
laser pulse with a duration (FWHM) of $\approx 140$ fs ($1.4$ ps).  
The calculations for
Fig.~\ref{fig:gauss} were performed using the 2D Markovian expression for the
rates in Eq.~(\ref{eq:rate}) and the renormalized precession frequencies
in Eq.~(\ref{eq:dw2d}).
It can be seen that the magnitude of the frequency renormalization
can reach values of several percent of the mean-field precession frequency 
and is negative for small values of $|\langle\mbf S\rangle|$.
For larger values of the impurity magnetization, the frequency renormalization
approaches a small positive value.
One could expect that the narrower electron distribution ($E_s=0.1$ meV) 
is closer to the $\delta$-like case than the wider distribution ($E_s=1$ meV)
and therefore the frequency renormalization should be more pronounced. 
However, it can be seen from Fig.~\ref{fig:gauss} that this is only the case
for very low values of $|\langle\mbf S\rangle|$ (below 0.01 in the case studied
here). 
For higher values of the impurity magnetization,
the relative frequency renormalization approaches the positive background 
much faster in the calculations with the narrower electron distribution.

Note that in order to
be able to measure or fit a precession frequency, at least one period 
of the oscillations should be visible before the spin polarization is decayed.
Thus, the minimal value of the impurity magnetization, 
where one can reasonable deduce a precession 
frequency from the time evolution of the spin polarization, is given by
$|\omega_M|\gtrsim \tau^{-1}_\perp$ which yields, for the parameters above,
$|\langle \mbf S\rangle|\gtrsim 0.01$.
Therefore, we find that short laser pulses with pulse durations of the order 
of 100 fs provide the most promising configuration for experiments to measure
the frequency renormalization.

\begin{figure}[t]
\includegraphics{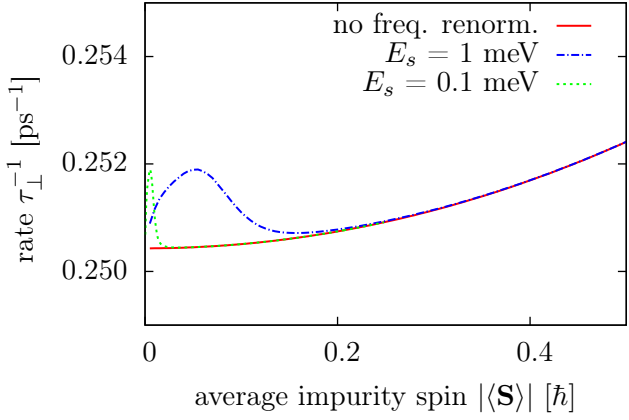}  
\caption{Rate $\tau_\perp^{-1}$
obtained from an exponential fit for Gaussian initial 
spectral electron distributions (cf. Fig.~\ref{fig:gauss}) as a function
of the average impurity spin with 
(blue dash-dotted line/green dotted line) and without (red solid line) 
accounting for a 
renormalization of the precession frequency. 
}
\label{fig:gauss_rate}
\end{figure}

Since the frequency renormalization depends on the kinetic energy and 
therefore the wave-vector of an electron, the question arises, whether 
this dependence leads to a dephasing of spins of electrons with different
$\k$-vectors. To address this question, we show in Fig.~\ref{fig:gauss_rate}
the value of the rate $\tau_\perp^{-1}$ obtained by an exponential fit
to the time evolution of the total carrier spin polarization, where the same
Gaussian initial electron distributions are used as in Fig.~\ref{fig:gauss}. 
It can be seen that calculations, where the correlation induced frequency 
renormalization is neglected, produce very similar decay rates as
calculations that account for this renormalization for most of the 
possible values of the impurity magnetization. Only in a regime where the
impurity spin is small we find 
a slightly larger value ($\lesssim 1\%$) of 
the rate at $|\langle\mbf S\rangle|\approx 0.05$ for $E_s=1$ meV and
$|\langle\mbf S\rangle|\approx 0.005$ for $E_s=0.1$ meV. 
This increasing decay 
is the consequence of the dephasing of electron spins due to the
$\k$-dependence of the frequency renormalization. Since the expression for
the rate in the Markov limit [cf. Eq.~(\ref{eq:rate})] and the 
frequency renormalization [cf. Eq.~(\ref{eq:dw2d})] depend on the 
same parameters, this dephasing mechanism is always accompanied by a
genuine spin transfer between impurities and carriers that is 
typically much faster than the dephasing itself.

\section{Correlation Energy}

Most studies on DMS which probe the energies of electrons in DMS
use the mean-field approximation\cite{FurdynaReview} and 
describe the effects of the impurity magnetization as a renormalization
of the electron g-factor which is known as the giant Zeeman effect\cite{CrookerTrions00}.
If, however, the build-up of carrier-impurity correlations is taken into
account, the mean $s$-$d$ exchange interaction energy $\langle H_{sd}\rangle$
will deviate from the mean-field value. The correlation energy can, 
in principle, have an impact on the thermodynamic properties of DMS which
could help, e.~g., in the description of the paramagnetic-ferromagnetic phase 
transition in GaMnAs.

Since the derivation of the
PESC-equations in Ref.~\onlinecite{PESC} 
required finding explicit expressions for the correlations, we can 
use this theory to get the correlation induced correction
$\langle H_{sd}^{cor}\rangle$
to the mean-field exchange interaction energy analytically:

\begin{align}
\langle H_{sd}\rangle&=
\frac\Jsd V\!\!\sum_{\substack{Inn'\\\sigma\sigma'\k\k'}}\!\!
\mbf S_{nn'}\mbf s_{\sigma\sigma'}
\langle c^\dagger_{\sigma\k}c_{\sigma'\k'}e^{i(\k'-\k)\mbf R_I}
\hat{P}^I_{nn'}\rangle=\nn&=:
\sum_{\k}\hbar\boldsymbol\omega_M \cdot \mbf s_{\k}+
\langle H_{sd}^{cor}\rangle 
\end{align}

Using the time-integral form of the correlations from 
Ref.~\onlinecite{PESC} we find:
\begin{widetext}
\begin{align}
\langle H_{sd}^{cor}\rangle&=
\frac\Jsd V\nMn \sum_{\k\k'}\sum_{\alpha=1}^3 Q^{\alpha\k'}_{\alpha\k}=
-\hbar\sum_{\k_1\k_2}\bigg\{ \Im\{ G_{\omega_{\k_2}}^{\omega_{\k_1}+\omega_M}\} 
\Big[\frac{b^+}2n^\down_{\k_2}-\frac{b^-}2n^\up_{\k_1}
-\frac{b^0}2(n^\up_{\k_1}n^\down_{\k_2}+
n^\up_{\k_2}n^\down_{\k_1})\Big]+ \nn&+
\Im\{ G_{\omega_{\k_2}}^{\omega_{\k_1}-\omega_M}\}
\Big[\frac{b^-}2n^\up_{\k_2}-\frac{b^+}2n^\down_{\k_1}
+\frac{b^0}2(n^\up_{\k_1}n^\down_{\k_2}+
n^\up_{\k_2}n^\down_{\k_1})\Big]+
\Im\{ G_{\omega_{\k_2}}^{\omega_{\k_1}}\}
\Big[\frac{b^\|}4\big((n^\up_{\k_2}+n^\down_{\k_2})
-(n^\up_{\k_1}+n^\down_{\k_1})\big)\Big]
\bigg\}
\label{eq:EcorFull}
\end{align}
\end{widetext}
To understand Eq.~(\ref{eq:EcorFull}) it is important to recall that
the correlations typically 
build up on the timescale of a few fs\cite{Preceedings},
while the spin-up and spin-down occupations change on a ps 
timescale\cite{Thurn:13_1,PESC}. Thus, $\langle H_{sd}^{cor}\rangle$ can
be interpreted as a quasi-equilibrium value of the correlation energy for
given values of adiabatically changing occupations $n^{\up/\down}_{\k_1}$.

As in the discussion of the frequency renormalization, we neglect terms of 
second order in the electron variables and apply the Markov approximation to
find for the two-dimensional case:

\begin{align}
\langle H_{sd}^{cor}\rangle\approx&
-\frac{\Jsd^2}{\hbar}\frac{\nMn}V\frac{Am^*}{2\pi\hbar}\sum_{\k_1}\nn&
\bigg\{ 
\ln\Big|\frac{\omega_{BZ}-(\omega_{\k_1}+\omega_M)}{\omega_{\k_1}+\omega_M}
\Big|b^- n^\up_{\k_1}+\nn&
+\ln\Big|\frac{\omega_{BZ}-(\omega_{\k_1}-\omega_M)}{\omega_{\k_1}-\omega_M}
\Big|b^+ n^\down_{\k_1}+\nn&
+\ln\Big|\frac{\omega_{BZ}-\omega_{\k_1}}{\omega_{\k_1}}\Big|
\frac{b^\|}2(n^\up_{\k_1}+n^\down_{\k_1})
\bigg\}
\label{eq:HcorMarkov}
\end{align}
The mathematical structure of the correlation energy 
$\langle H_{sd}^{cor}\rangle$ in 
Eq.~(\ref{eq:HcorMarkov}) is very similar to that of the
frequency renormalization in Eq.~(\ref{eq:freqrengen}).
To see this relation, it is helpful to express the 
occupations $n^{\up/\down}_{\k_1}$ of the spin-up and spin-down 
subbands in terms of the occupation 
$n_{\k_1}$ of both bands and the spin component $s^\|_{\k_1}$ 
parallel to the impurity magnetization (quantization axis) via
\begin{align}
n^{\up/\down}_{\k_1}=\frac{n_{\k_1}}2\pm s^\|_{\k_1}.
\end{align}
As it is common for spin-dependend single particle energies like 
the Dresselhaus-\cite{Dresselhaus} or Rashba-terms\cite{Rashba},
one could expect that the spin-dependend part of the correlation energy
can be written as an effective magnetic field in which the electron
spins precess. This additional precession movement could be made responsible
for the frequency renormalization discussed above. However, although 
the corresponding effective field due to the correlation energy has the 
same form as the frequency renormalization, it is larger by a factor of 2.
We attribute this to the fact that
the correlation energy is not an average over a
Hermitian single particle operator, but comprises multiparticle effects, 
where the naive identification of an effective magnetic field can lead
to incorrect predictions. 

A particularly interesting and transparent case is that where the 
impurity magentization $\langle\mbf S\rangle$ vanishes. Then,
the correlation energy takes the form
\begin{align}
\langle H_{sd}^{cor}\rangle&=
-\frac{\Jsd^2}{\hbar}\frac{\nMn}V \langle S^2\rangle\frac{Am^*}{2\pi\hbar}
\sum_{\k_1}\ln\Big|\frac{\omega_{BZ}-\omega_{\k_1}}{\omega_{\k_1}}\Big|
n_{\k_1}
\label{eq:HcorS0}
\end{align}
Thus, for $\langle \mbf S\rangle=0$, we find a logarithmic divergence of
the correlation energy with respect to $\omega_{BZ}\to\infty$ and 
$\omega_{\k_1}\to 0$. In both limits, the correlation energy is negative
and independent of the sign of the coupling constant $\Jsd$. 
The negative correlation energy suggests a formation of 
correlated carrier-impurity states, i.~e., magnetic polarons.
Such states have also been predicted in GaMnAs\cite{BMPM12}, 
where the correlation energy stems from an attractive Coulomb potential
due to the charged state of the Mn ions which acts, e.~g., as acceptors, if
they substitute Ga atoms in the GaAs crystal lattice.
An additional contribution to the correlation energy originating from 
the spin-dependent $s$-$d$ interaction could enhance the formation 
of the magnetic polarons in GaMnAs.
Furthermore, our finding is consistent with the fact that also in the 
solution of the Kondo problem, the ground state is strongly 
correlated\cite{Hewson}.

Again, the divergence at $\omega_{\k_1}\to 0$ in Eq.~(\ref{eq:HcorS0})
is integrable so that the total correlation energy always assumes finite values.
To estimate the magnitude of the correlation energy we consider 
the case where $\langle \mbf S\rangle=0$ and the spectral electron distribution
is given by a Gaussian centered at the band edge with standard deviation 
$E_s=1$ meV. For the parameters of the Cd$_{0.93}$Mn$_{0.07}$Te 
quantum well discussed above, we find from Eq.~(\ref{eq:HcorS0}) a value 
of $\langle H_{sd}^{cor}\rangle\approx$-1.8 $\mu$eV  per electron. Thus,
the correlations can be destroyed by thermal fluctuations when the temperature
$T$ exceeds $20$ mK.

\section{Conclusion}
A microscopic quantum kinetic theory is employed to describe the 
spin dynamics of carriers and magnetic impurities in diluted magnetic
semiconductors (DMS) accounting also for the dynamics of the
carrier-impurity correlations. The role of the correlations is examined 
to shed light into the controversy
about their importance: While some authors assume that the Kondo physics
due to carrier-impurity correlation is of minor importance\cite{DasSarma03},
others\cite{Morandi10} find divergences in a perturbative treatment of
the spin dynamics in DMS, similarly to the appearance of divergences found
in the metallic Kondo-effect\cite{Kondo}.
In the present study, we find that the correlations, besides mediating the
spin transfer between carriers and impurities, are also responsible for
a renormalization of the precession frequency of up to a few percent in 
two-dimensional structures.  
Since for these values, the relative frequency renormalization is negative,
the precession frequency of the electron spin is reduced. 

In order to experimentally probe the correlation induced frequency 
renormalization, the spectral features of the laser pulse have to be
precisely controlled.
Furthermore, it was reported\cite{GPF} that an antiferromagnetic 
impurity-impurity interaction influences the thermal equilibrium value
of the Mn magnetization, which in turn changes the measured 
electron spin precession frequency. Therefore, it is common to 
introduce a fitting parameter $T_{0}$ and describe the equilibrium Mn 
magnetization by a Brillouin function with effective temperature 
$T_{eff}=T_0+T$, where $T$ is the temperature of the sample.
This complicates the identification of
correlation induced changes in the precession frequency. 
To distinguish both effects it is useful that in addition to the
dependence on the spectral position and shape of the exciting pulse,
the relative frequency renormalization due to the correlations is 
independent of the impurity density, while the impurity-impurity interaction
depends on the mean distance between the impurity ions and is not influenced
by the excitation conditions.
Because of this and from the different parameters entering the prefactor of 
the frequency renormalization, we find
that the most promising samples for experimentally accessing the correlation
induced frequency renormalization are very narrow quantum 
wells with large effective masses and a large coupling constant $\Jsd$ 
while the impurity concentration should be relatively low.
Also, we find that the spectral properties of ultrashort pulses with
durations in the 100 fs range suit this purpose. 

Although the $\k$-dependence of the frequency renormalization can in principle
lead to a dephasing of carrier spins, the spin transfer from the carriers to
the impurities is usually much faster, so that this dephasing mechanism yields
only very small corrections to the total decay of the carrier spin.

By comparing the results
of the quantum kinetic dynamics including the time evolution of correlations 
with the Markov limit, where the correlations can be expressed in terms of
carrier and impurity variables alone, we find that simple Markovian expressions
for spin transfer rate and frequency renormalization reproduce the results 
of the quantum kinetic treatment very well.
Although logarithmic divergences are found for some values of carrier 
wave-vectors as well as for bandwidth $\to\infty$, similar to the metallic 
Kondo effect, 
the total frequency renormalization, which is obtained by integrating 
with a smooth spectral carrier-distribution, is always finite.
Also, the divergences only appear in the Markov limit, while the effects
induced by the correlations are analytic by construction when a finite
memory is accounted for.

The explicit expressions for the correlations in the Markov limit also allow
to find an equilibrium value for the correlation energy in terms of carrier
and impurity variables. The form of the correlation energy is similar to
the expression for the frequency renormalization and hints towards the
appearance of correlated carrier-impurity states for low 
temperatures, independent of the sign of the coupling constant.
However, the required temperatures are lower than the values that
are typically considered in experimental setups probing DMS. Thus, in these
cases, the correlation energy is not expected to influence the 
thermodynamic properties of DMS strongly.

Although many similarities between the DMS systems and the Kondo problem 
can be found, a one-to-one correspondence can not be established since
the density of states in semiconductors is much smaller than in metals. 
Hence, the assumption that the number of magnetic impurities exceed the number
of quasi-free carriers, that is valid for DMS and has been used in the 
derivation of the equations of motion on which the present study is 
based\cite{PESC}, is opposite to the situation in usual metallic Kondo-systems.
Furthermore, the large number of carriers in the metal can be expected to
lead to a fast carrier spin relaxation due to Coulomb scattering.
However, the approach of finding a quasi-equilibrium value for the
correlation energy from a quantum kinetic theory could be extended and
might help to understand the Kondo physics from a new point of view.
Since we find that the correlation effects on the spin dynamics and the
correlation energies are linear in the density of states, our theory 
indeed agrees with the finding of strongly correlated states for 
low temperatures in metallic Kondo systems. 

\begin{acknowledgments}
We gratefully acknowledge the financial support of the 
Deutsche Forschungsgemeinschaft through grant No.\ AX17/9-1,
from the Universidad de Buenos Aires, 
project UBACyT 2011-2014 No. 20020100100741, 
and from CONICET, project PIP 11220110100091.
\end{acknowledgments}
\bibliography{alle}

\end{document}